%% file: sigmac_width.tex
\documentclass{elsart}
\input{sc_width_setup.tex}

\begin{document}

\begin{frontmatter}
\title{\boldmath Measurement of Natural Widths of \scz and \scd Baryons}

\date{\today}

\input{plb_list.tex}

%\pacs{14.20.Lq,13.30Eg}
 
\begin{abstract}
  In this paper we present a measurement of the natural widths of
  \scz and \scd. Using data from the FOCUS experiment, we find $\gscz = \sczga
  \sczst \sczsy$~\mevcc and $\gscd = \scdga \scdst \scdsy$~\mevcc. The first
  errors are statistical, the second systematic. These results are obtained with
  a sample of \scznum \lcpim decays and \scdnum \lcpip decays. These results are
  compared with recent theoretical
  predictions. 
  
  PACS numbers: 14.20.Lq 13.30Eg 
\end{abstract}
\end{frontmatter}

%\maketitle

%%% Body begins here

The natural widths of the excited charmed baryons have become experimentally
accessible~\cite{Albrecht:1997qa,Edwards:1994ar,Brandenburg:1997jc} only in the
last several years. The widths of most of the known excited charmed baryons are
poorly  determined with either no existing measurement or with an upper
limit;\footnote{During final preparation of this manuscript, we became aware of
a measurement of \scz and \scd intrinsic widths by the CLEO
collaboration~\cite{Artuso:cleo_width}.} most of the lower mass excited states
appear to have widths comparable to or less than the resolution of current
experiments.  Combined with relatively low statistics, this has made such
measurements challenging.  Direct measurements of the \sigc widths are
important since most of the current theoretical models predict the widths of
charmed baryons by extrapolating from the hyperon widths.  Accurate
measurements of the \sigc  widths will enable more accurate predictions of the
widths of these and other excited charm states and test the physics underlying
these models.

In this paper, we use data from the FOCUS experiment to obtain measurements
of \gscz and \gscd. The \scz and \scd mass differences with respect to the \lc
were presented in an earlier paper~\cite{Link:2000qs}. FOCUS is an upgraded
version of FNAL-E687~\cite{Frabetti:1992au} which collected data
using the Wideband photon beamline during the 1996--1997 Fermilab fixed-target
run.  The FOCUS
experiment utilizes a forward multiparticle spectrometer to study charmed
particles produced by the interaction of high energy photons  ($\langle E
\rangle \approx 180~\gev$) with a segmented BeO target.  

Charged particles are tracked within the spectrometer by two silicon
microvertex detector systems. One system is interleaved with the target
segments; the other is downstream of the target region. These detectors provide
excellent separation of the production and decay vertices.  Further downstream,
charged particles are tracked and momentum analyzed by a system of five
multiwire proportional chambers and two dipole magnets with opposite polarity.
Three multicell threshold \cer detectors are used to discriminate among
electrons, pions, kaons, and protons~\cite{Link:2001pg}.

The \sigc candidates are reconstructed via the decay chain $\sigc \to \lc
\pion^{\pm}$, with  \lc candidates found in four decay
modes,\footnote{Throughout, charge conjugate states are implied.}
$\lc \to$ $\proton \kminus \piplus$, $\proton \kshort$, $\lb \piplus$, and $\lb
\piplus \piminus \piplus$.  The reconstruction of \kshort's and \lb's is
described elsewhere~\cite{Link:2001dj}. Due to
topological differences and varying levels of background,  the values of the
analysis cuts for each of the four \lc decay modes vary. 
A decay, or secondary, vertex is formed from
selected reconstructed tracks; the momentum vector of this charm candidate
is then used as a seed to intersect other reconstructed tracks in the event
to find the production, or primary, vertex~\cite{Frabetti:1992au}.  We impose a minimum detachment cut
which requires that the measured separation of these two vertices divided by the
error on that measurement be greater than our cut (typically 3--5). We also
ensure that both vertices are well formed by requiring a confidence level
greater than 1\% on the fit to each vertex.  To remove longer lived
charm backgrounds we also require the reconstructed proper lifetime to be less
than some amount, typically 4--5 times the mean \lc lifetime. Finally, momentum
and particle identification cuts are applied to each decay mode.  Invariant mass plots
for each of the decay modes are shown in \figref{lc_modes}.  

\begin{figure}
 \begin{center}
  \includegraphics[width=8.4cm,bb=0 220 590 770]{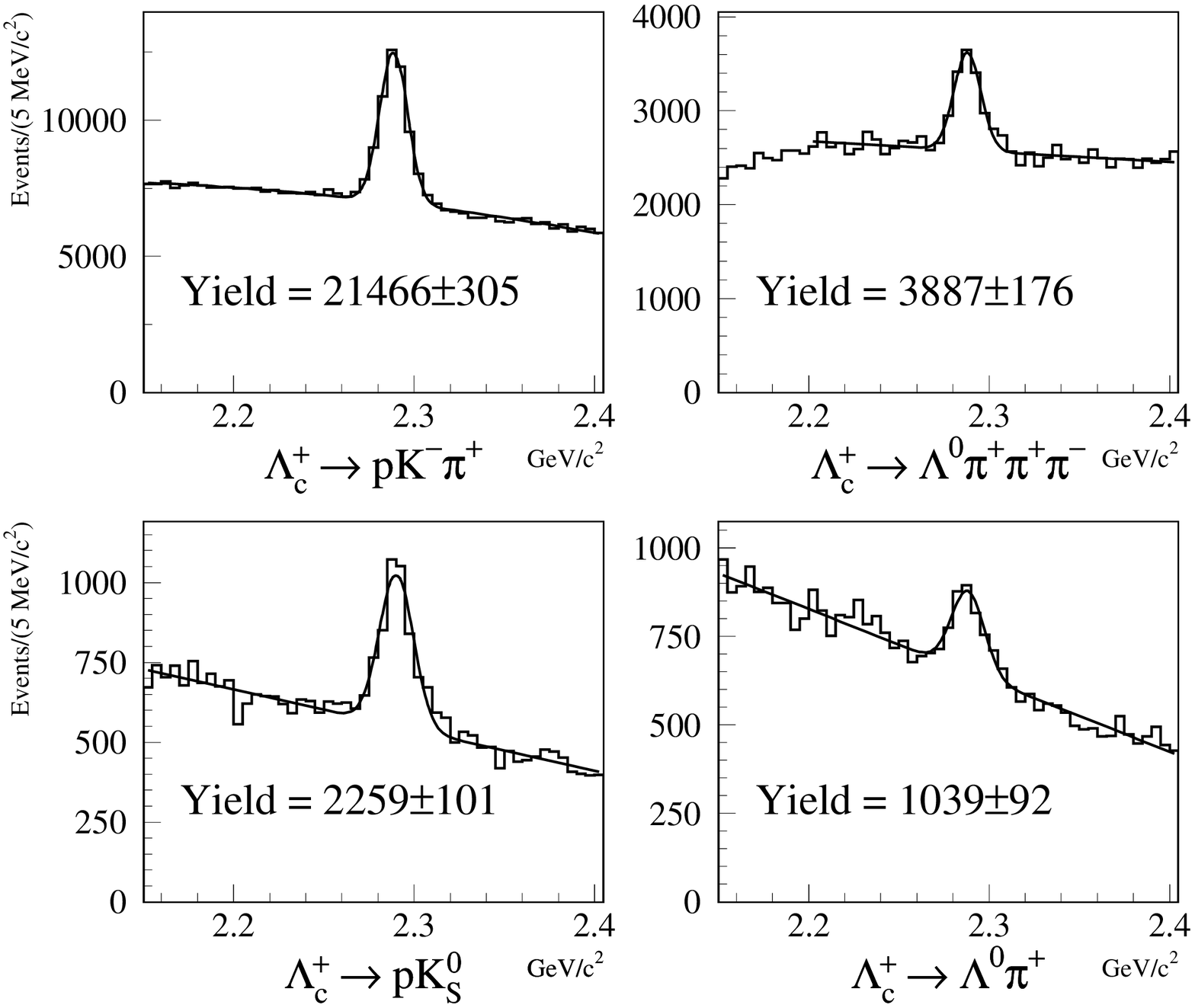}\\
  \includegraphics[width=4.2cm,bb=0 510 295 720]{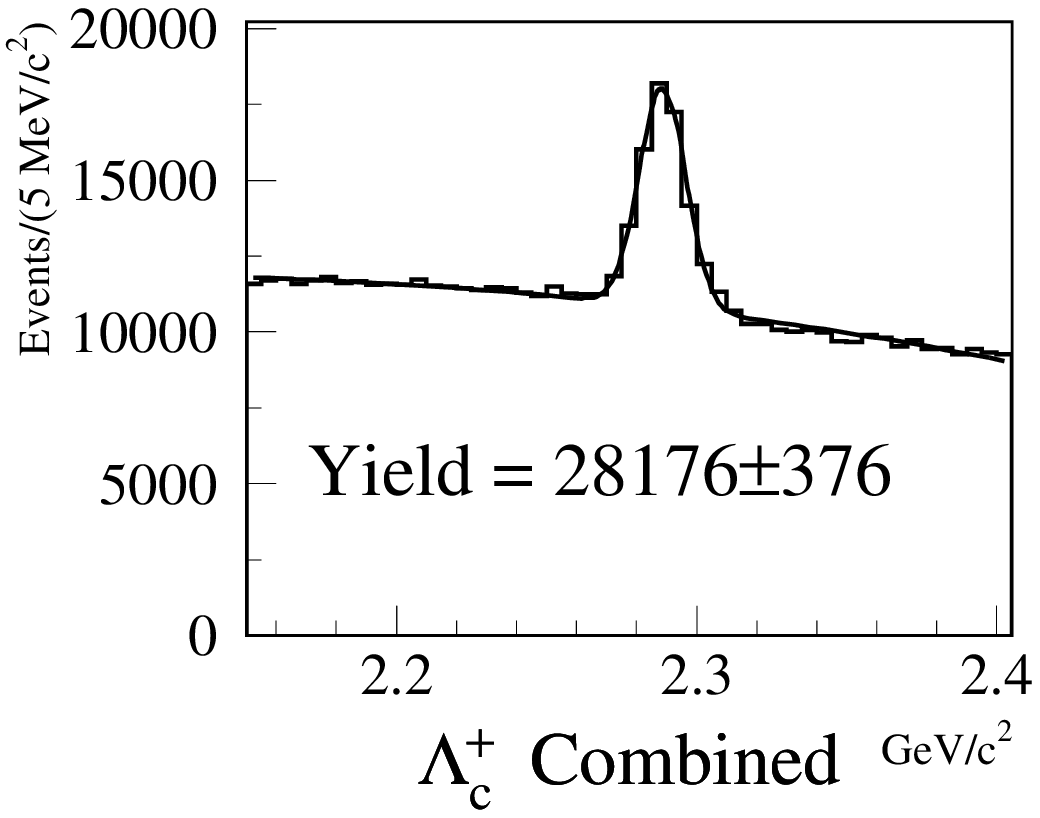}\\
 \end{center}
 \caption{Invariant mass distributions for \lc candidates. Shown are the four
 \lc decay modes  used in the analysis and, on the bottom, the combined invariant
 mass plot for all four decay modes.}
 \figlabel{lc_modes}
\end{figure}

The \sigc candidates are reconstructed by combining the \lc candidates within
approximately $2\, \sigma$ of the mean \lc mass with a charged
pion.\footnote{Referred to as a ``soft pion'' since it is usually a low
momentum particle.}   The vertex formed by the \lc candidate, the pion
candidate, and at least one other track must have a confidence level greater
than 1\%. \cer identification on the soft pion requires
that the pion hypothesis is not heavily disfavored with respect to any other
identification hypothesis.  

To remove systematic effects due to the reconstruction of the \lc mass, we
compute and plot the invariant mass differences ($\Delta M = M(\lc\pi^\pm) -
M(\lc)$).  Because most of the uncertainty in the mass difference arises due to
multiple scattering of the soft pion, we improve the measurement of the soft
pion momentum as follows. The primary vertex is refit without the soft pion and
the soft pion is  constrained to originate from this new primary vertex.  If
the confidence level of this constraint is less than 1\%, the candidate is
discarded. The measured \lc momentum and mass are combined with the constrained
pion momentum and known mass to form $M(\lc \pi^\pm)$.  The computed \lc mass
is subtracted to obtain the invariant mass difference, $\Delta M$.  The
reconstructed mass difference distributions are shown in \figref{sigmac}.   

\begin{figure}
 \begin{center}
  \includegraphics[width=8.4cm,bb=3 352 590 735]{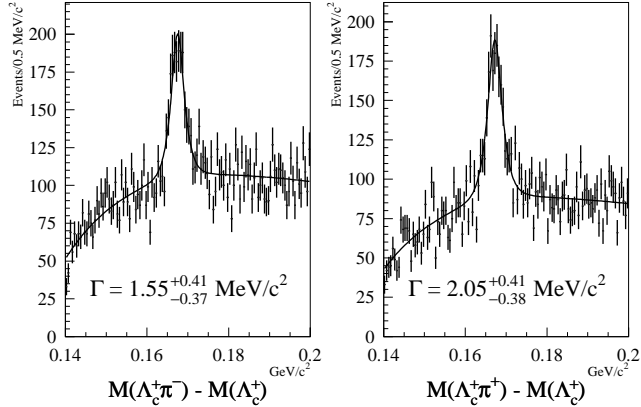}
 \end{center}
 \caption{Invariant mass differences for \scz (left) and \scd (right)
 candidates. The fit functions, over their range of 140--200~\mevcc, are shown
 as are
 the widths and statistical errors.}
 \figlabel{sigmac}
\end{figure}

%\section{Measurement of $\Gamma(\sigc)$}

To measure $\Gamma(\sigc)$, we fit the \sigc signal distributions with a
relativistic constant width \bw function convoluted with a parameterization of
the experimental resolution derived from \mc.  The experimental resolution is
determined by generating \mc events with $\gsc = 0$ and fitting the resulting
\sigc distributions to a sum of two Gaussians,
$G(x;\text{Yield},\sigma,\bar{x})$, with the same central value $\bar{x}$: 
\begin{equation}
%\begin{split}
%  A\left((1-f) \exp \left[-\frac{1}{2} \left(\frac{\Delta M -
%  m_{\sigc}}{\sigma_1}\right)^2 \right] \right.
%  + \\ \left. f \exp \left[-\frac{1}{2} \left(\frac{\Delta M -
%  m_{\sigc}}{\sigma_2}\right)^2\right] \right)
%\end{split}
G \left( \Delta M; (1-f)Y,\sigma_1, M_{\text{fit}}  \right) + 
         G \left( \Delta M ; fY,\sigma_2, M_{\text{fit}}\right) \, . 
  \eqnlabel{double_g}
\end{equation} 
We use a background function described by
\begin{equation}
 N(1 + \alpha(\Delta M-m_\pi)\Delta M^{\beta})
 \eqnlabel{free_bg}
\end{equation}
where $m_\pi$ is the $\pi^\pm$ mass.  $N$, $\alpha$, and $\beta$ are allowed to
vary.  Three parameters of interest are extracted from the fit: $\sigma_1$,
$\sigma_2$, and $f = Y_2/Y$ which describe, respectively, the resolution of the
narrow and wide portions of the resolution function and the ratio of the second
yield to the total. The extracted parameters are shown in \tabref{sigmac_res}. $M_{\text{fit}}$ is the $\sigc-\lc$ mass difference.

\begin{table}
\centering
\caption[\sigc resolution parameters]{\sigc resolution parameters calculated in
\mc.}
%\begin{ruledtabular}
\begin{tabular}{|l|c|c|c|}\hline
 &
$\sigma_1$ &
$\sigma_2$ &
$f$  \\ 
\multicolumn{1}{|c|}{State} &
(\mevcc) &
(\mevcc) &
$\text{Yield}_2/(\text{Yield}_1+\text{Yield}_2$)  \\ \hline
\scz & $0.84\pm0.03$ & $1.93\pm0.05$ & $0.53\pm0.03$ \\
\scd & $0.83\pm0.03$ & $1.92\pm0.05$ & $0.56\pm0.03$ \\
\hline
\end{tabular}
%\end{ruledtabular}
\tablabel{sigmac_res}
\end{table}

We determine the shape parameters ($\alpha$ and $\beta$) of the background
function (\eqnref{free_bg}) by fitting the sum of the $M(\lc\piplus)$ and the
$M(\lc\piminus)$ data distributions to \eqnref{free_bg} and a convoluted \bw
signal shape. The values of $\alpha$ and $\beta$ are fixed to these values when
fitting the individual distributions. 

While the fits for \scz and \scd distributions to a \bw convoluted with the
resolution function use a total of nine parameters, only four of these are free
parameters (yield, $M(\sigc-\lc)$, \gsc, and $N$, the background
normalization). The parameters $\alpha$, $\beta$, $\sigma_1$, $\sigma_2$, and
$f$ are previously determined and are fixed.  Performing the fit as
described above we obtain values of $\gscz = \sczga \sczst$~\mevcc and $\gscd =
\scdga \scdst$~\mevcc with yields of \sczyld and \scdyld respectively. The
extracted mass differences are consistent with our previous
measurement~\cite{Link:2000qs}.

%\section{Systematic studies}

The largest systematic uncertainties in this measurement arise from our
imperfect knowledge of the experimental mass resolution. To better understand our
experimental resolution in a kinematically similar situation, we have used the
decay $\dsp \to \dzero \piplus$ as a benchmark. The \dsp is much closer to
decay threshold and should expose any problems with our simulation of the
experimental resolution. We compare the resolution of classes of events seen in
the data with that seen in the simulation. The \mc  uses an input \dsp natural
width of 96~\kevcc, the value from a recent
CLEO~\cite{Ahmed:2001xc,Anastassov:2001cw} report. For both \mc and data we fit
the distributions to the sum of \eqnref{double_g} and \eqnref{free_bg}. No
value of $\Gamma(\dsp)$ is determined by or included in the fit. (Since
$\Gamma(\dsp) \ll \sigma(\Delta M(\dsp))$, the precise value of $\Gamma(\dsp)$
influences the fit results only slightly. The \bw modifications to the line
shape are negligible and are safely ignored.) By studying any discrepancies in
the \dsp case, we understand our experimental resolution better  for the \sigc.

We find that for the \dsp, our overall experimental resolution matches that
predicted by the \mc quite well and find no definitive evidence for any
discrepancy except for $\sigma_1$ which is about 10\% less in \mc. The values
of the fits to the \dsp line shape for data and \mc are shown in
\tabref{dstar_res}. To be conservative, we test the effects of varying the
experimental resolution by the maximum allowed by our \dsp studies. In our
description of the predicted resolution we independently vary the three
parameters $\sigma_1$, $\sigma_2$, and $f$ by $\pm 10\%$, $\pm 20\%$, and $\pm
15\%$ respectively. These changes induce variations of approximately $\mp
0.20$, $\mp 0.25$, and $\mp 0.20$~\mevcc in $\Gamma(\sigc)$, respectively. 

\begin{table}
\centering
\caption{Comparison of \dsp line shape parameters for data and \mc. Statistical
errors only.}
%\begin{ruledtabular}
\begin{tabular}{|l|c|c|c|}\hline
 &
$\sigma_1$ &
$\sigma_2$ &
 \\ 
\multicolumn{1}{|c|}{State} & (\mevcc) & (\mevcc) & $f$   \\ \hline
\dsp Data & $0.66\pm0.00$ & $3.5\pm0.2$ & $0.05\pm0.02$ \\
\dsp MC   & $0.61\pm0.00$ & $2.8\pm0.5$ & $0.05\pm0.02$ \\
\hline
\end{tabular}
%\end{ruledtabular}
\tablabel{dstar_res}
\end{table}

In addition, we have studied the resolution function for the \dsp as a function
of target configurations, the production target segment, decay modes,  and the
detachment cut. The resolution is clearly dependent on the production target
segment and the several target configurations used in FOCUS. The dependence of
the resolution on these two variables is well predicted by the \mc simulation
and the distribution of \sigc events as a function of these variables is also
well matched by the \mc. The resolution function also depends somewhat on the
kinematic variables, most strongly on the momentum of the excited state. As an
example, in \tabref{dstar_mom_comp},  we show the variation of $\sigma_1$ for
the \dsp in four bins of momentum. The data and \mc disagree by less than 10\%
as shown in \tabref{dstar_res}, but the disagreement is constant vs. \dsp
momentum. In \figref{mom_comp}, we show that the \sigc momentum distribution is
well modeled by the \mc. Other variations in resolution are found as a
function of soft pion momentum and charm and soft pion directions, with a
weaker dependence on these kinematic variables.

\begin{table}
\centering
\caption{Comparison of \dsp core resolution, $\sigma_1$ (\mevcc), vs. momentum.}
%\begin{ruledtabular}
\begin{tabular}{|l|c|c|c|c|}\hline
 $p(\dsp)$~\gevc & $<60$ & 60--80 & 80--100 & $>100$ \\  \hline
\dsp Data & $0.61\pm0.01$ & $0.63\pm0.01$ & $0.65\pm0.01$ & $0.70\pm0.01$  \\
\dsp MC   & $0.57\pm0.00$ & $0.59\pm0.00$ & $0.62\pm0.00$ & $0.66\pm0.00$  \\
\hline
\end{tabular}
%\end{ruledtabular}
\tablabel{dstar_mom_comp}
\end{table}

\begin{figure}
 \begin{center}
  \includegraphics[width=6.4cm,bb=0 480 300 735]{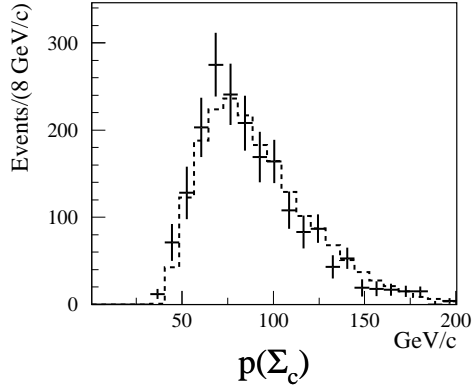}
 \end{center}
 \caption{Momentum spectrum of \scz and \scd (summed) for data (points) and \mc
 (dashed curve). Both \mc and data are sideband subtracted.}
 \figlabel{mom_comp}
\end{figure}

%We have studied the mean reconstructed $\Delta M$ as a function of the
%\sigc decay mode, since any discrepancies might increase the apparent width. We
%find no evidence for such an effect.

We also perform two additional tests of our fitting and reconstruction method.
In the first, we produce a large number of distributions which are
statistically similar to the data. These distributions are then fit with the
same methods used for the real data. The extracted fit parameters are
compared with the known input parameters. From this source we estimate a
maximum bias in the fitting technique of $0.05$~\mevcc and confirm the validity
of our statistical errors.

In the second test, we generate \mc events with a natural width which is
comparable to what is measured and measure the width of these events using the
same methods. From this study we find no evidence for a systematic error.

Adding all the systematic errors in quadrature, we find a total systematic error of
$\pm \scsy$~\mevcc for each \gsc measurement. This gives final values of $\gscz
= \sczga \sczst \sczsy$~\mevcc and $\gscd = \scdga \scdst \scdsy$~\mevcc.

%\section{Theory}  %% Theory

There are several recent theoretical predictions for the widths of the
\sigc 
states \cite{Ivanov:1998qe,Ivanov:1999bk,Tawfiq:1998nk,Huang:1995ke,Pirjol:1997nh,Rosner:1995yu};
all predict $\Gamma(\sigc)$ in the range 1--3~\mevcc.  Several different
theoretical models are used, including the Relativistic Three Quark Model
(RTQM), Heavy Hadron Chiral Perturbation Theory (HHCPT), and the Light Front
Quark Model (LFQM).  

These models predict partial widths, but since \lcpi is the only allowed strong or
electromagnetic decay mode for the states under study, we safely
take the partial width as the total width for each state.  The resulting
predictions from the models are shown in \tabref{theor_pred}. Our measurements
are not of sufficient precision to strongly favor any one of these models over
another.

\begin{table}
\centering
\caption[Predicted \sigc widths from various theoretical models]{Predicted \sigc widths from various theoretical models.  All units are \mevcc.}
%\begin{ruledtabular}
\begin{tabular}{|l|l|l|l|}\hline
\multicolumn{1}{|c|}{Author} &
\multicolumn{1}{c|}{$\Gamma(\scz)$} &
\multicolumn{1}{c|}{$\Gamma(\scd)$} &
\multicolumn{1}{c|}{Method }  \\ \hline
Ivanov, \etal~\cite{Ivanov:1998qe,Ivanov:1999bk} & 
                              $2.65 \pm 0.19$ & $2.85 \pm 0.19$ & RTQM \\
Tawfig, \etal~\cite{Tawfiq:1998nk} & $1.57$ & $1.64$ & LFQM   \\
Huang, \etal~\cite{Huang:1995ke} &   $2.4$ & $2.5$ & HHCPT   \\
Pirjol, \etal~\cite{Pirjol:1997nh} & 1.0--3.0 & 1.1--3.1 & HHCPT \\
Rosner~\cite{Rosner:1995yu} & $1.32 \pm 0.04$ & $1.32 \pm 0.04$ & Ratios   \\
\hline
\end{tabular}
%\end{ruledtabular}
\tablabel{theor_pred}
\end{table}

%\section{Conclusion}

In conclusion, we present measurements of the natural widths of
the \scz and \scd excited charmed baryons using data from the FOCUS experiment. We find $\gscz = \sczga \sczst
\sczsy$~\mevcc and $\gscd = \scdga \scdst \scdsy$~\mevcc which are consistent
with most current theoretical predictions.

%\begin{acknowledgments}
\input{acknowledgements}

%\end{acknowledgments}

\bibliographystyle{elsart-num}
\bibliography{sigmac_width_paper}

\end{document}

%% file: sc_width_setup.tex
\usepackage{graphicx}
\usepackage{ifthen}
\usepackage{xspace}

\usepackage{amsmath}

\newcommand{\gsc}{\ensuremath{\Gamma(\sigc)}\xspace}
\newcommand{\gscz}{\ensuremath{\Gamma(\scz)}\xspace}
\newcommand{\gscd}{\ensuremath{\Gamma(\scd)}\xspace}

\newcommand{\scsy}{0.38}

\newcommand{\sczga}{\ensuremath{1.55}\xspace}
\newcommand{\sczst}{\ensuremath{^{+0.41}_{-0.37}}\xspace}
\newcommand{\sczsy}{\ensuremath{\pm \scsy}\xspace}

\newcommand{\scdga}{\ensuremath{2.05}\xspace}
\newcommand{\scdst}{\ensuremath{^{+0.41}_{-0.38}}\xspace}
\newcommand{\scdsy}{\ensuremath{\pm \scsy}\xspace}

\newcommand{\scznum}{\ensuremath{913}\xspace}
\newcommand{\scdnum}{\ensuremath{1110}\xspace}
\newcommand{\sczyld}{\ensuremath{\scznum \pm 77}\xspace}
\newcommand{\scdyld}{\ensuremath{\scdnum \pm 83}\xspace}

\newcommand{\cer}{\v{C}erenkov\xspace}

\newcommand{\mc}{Monte Carlo\xspace}
\newcommand{\bw}{Breit--Wigner\xspace}

\newcommand{\mathsl}[1]{\mbox{\textsl{#1}}}

\newcommand{\meson}[1]{\ensuremath{\mathsl{#1}}}
\newcommand{\baryon}[1]{\ensuremath{\mathsl{#1}}}

\newcommand{\proton}{\baryon{p}\xspace}

\newcommand{\pion}{\ensuremath{\pi}\xspace}
\newcommand{\piplus}{\ensuremath{\pion^+}\xspace}
\newcommand{\piminus}{\ensuremath{\pion^-}\xspace}

\newcommand{\kaon}{\ensuremath{\meson{K}}\xspace}
\newcommand{\kminus}{\ensuremath{\kaon^-}\xspace}

\newcommand{\kshort}{\ensuremath{\kaon_S^0}\xspace}

\newcommand{\lb}{\ensuremath{\Lambda^0}\xspace}

\newcommand{\dmeson}{\ensuremath{\meson{D}}\xspace}
\newcommand{\dzero}{\ensuremath{\dmeson^{0}}\xspace}
\newcommand{\dsp}{\ensuremath{\dmeson^{*+}}\xspace}

\newcommand{\lc}{\ensuremath{\Lambda_c^+}\xspace}

\newcommand{\sigc}{\ensuremath{\Sigma_c}\xspace}
\newcommand{\scd}{\ensuremath{\Sigma_c^{++}}\xspace}

\newcommand{\scz}{\ensuremath{\Sigma_c^{0}}\xspace}

\newcommand{\lcpi}{\ensuremath{\sigc \to \lc \pi}\xspace}
\newcommand{\lcpip}{\ensuremath{\scd \to \lc \piplus}\xspace}

\newcommand{\lcpim}{\ensuremath{\scz \to \lc \piminus}\xspace}

\newcommand{\kevcc}{\ensuremath{\mathrm{keV}/c^2}\xspace}

\newcommand{\mevcc}{\ensuremath{\mathrm{MeV}/c^2}\xspace}

\newcommand{\gevc}{\ensuremath{\mathrm{GeV}/c}\xspace}
\newcommand{\gev}{\ensuremath{\mathrm{GeV}}\xspace}

\newcommand{\eqnref}[1]{Eq.~(\ref{eqn:#1})}
\newcommand{\figref}[1]{Fig.~\ref{fig:#1}}
\newcommand{\tabref}[1]{Table~\ref{tab:#1}}

\newcommand{\figlabel}[1]{\label{fig:#1}}
\newcommand{\eqnlabel}[1]{\label{eqn:#1}}
\newcommand{\tablabel}[1]{\label{tab:#1}}

%% file: plb_list.tex
The FOCUS Collaboration

%%%%%%% Do not change authors here.  Use the database!
\author[ucd]{J.~M.~Link}
\author[ucd]{M.~Reyes}
\author[ucd]{P.~M.~Yager}
\author[cbpf]{J.~C.~Anjos}
\author[cbpf]{I.~Bediaga}
\author[cbpf]{C.~G\"obel}
\author[cbpf]{J.~Magnin}
\author[cbpf]{A.~Massafferri}
\author[cbpf]{J.~M.~de~Miranda}
\author[cbpf]{I.~M.~Pepe}
% ...

\address[ucd]{University of California, Davis, CA 95616} 
\address[cbpf]{Centro Brasileiro de Pesquisas F\'\i sicas, Rio de Janeiro, RJ, Brasil} 
\address[cinv]{CINVESTAV, 07000 M\'exico City, DF, Mexico} 
\address[cu]{University of Colorado, Boulder, CO 80309} 
\address[fnal]{Fermi National Accelerator Laboratory, Batavia, IL 60510} 
\address[fras]{Laboratori Nazionali di Frascati dell'INFN, Frascati, Italy I-00044} 
\address[ui]{University of Illinois, Urbana-Champaign, IL 61801} 
\address[iu]{Indiana University, Bloomington, IN 47405} 
\address[korea]{Korea University, Seoul, Korea 136-701} 
\address[milan]{INFN and University of Milano, Milano, Italy} 
\address[nc]{University of North Carolina, Asheville, NC 28804} 
\address[pavia]{Dipartimento di Fisica Nucleare e Teorica and INFN, Pavia, Italy} 
\address[pr]{University of Puerto Rico, Mayaguez, PR 00681} 
\address[sc]{University of South Carolina, Columbia, SC 29208} 
\address[ut]{University of Tennessee, Knoxville, TN 37996} 
\address[vu]{Vanderbilt University, Nashville, TN 37235} 
\address[wisc]{University of Wisconsin, Madison, WI 53706}

\endnote{\small See http://www-focus.fnal.gov/authors.html for
additional author information}

%% file: acknowledgements.tex
We wish to acknowledge the assistance of the staffs of Fermi National
Accelerator Laboratory, the INFN of Italy, and the physics departments of the
collaborating institutions. This research was supported in part by the U.~S.
National Science Foundation, the U.~S. Department of Energy, the Italian
Istituto Nazionale di Fisica Nucleare and Ministero dell'Universit\`a e della
Ricerca Scientifica e Tecnologica, the Brazilian Conselho Nacional de
Desenvolvimento Cient\'{\i}fico e Tecnol\'ogico, CONACyT-M\'exico, the Korean
Ministry of Education, and the Korean Science and Engineering Foundation.